\documentstyle[epsf,rotate]{article}
 
\catcode`\@=11
\long\def\@makefntext#1{ 
\protect\noindent \hbox to 3.2pt {\hskip-.9pt
$^{{\eightrm\@thefnmark}}$\hfil}#1\hfill} 
 
\def\thefootnote{\fnsymbol{footnote}}
 \def\@makefnmark{\hbox to 0pt{$^{\@thefnmark}$\hss}}  
 
\def\ps@myheadings{\let\@mkboth\@gobbletwo
\def\@oddhead{\hbox{} 
\rightmark\hfil\eightrm\thepage}
\def\@oddfoot{}\def\@evenhead{\eightrm\thepage\hfil 
\leftmark\hbox{}}\def\@evenfoot{}
\def\sectionmark##1{}\def\subsectionmark##1{}}
 
\oddsidemargin=\evensidemargin
\renewcommand{\thefootnote}{\fnsymbol{footnote}}

\newcounter{sectionc}\newcounter{subsectionc}\newcounter{subsubsectionc}
\renewcommand{\section}[1] {\vspace{12pt}\addtocounter{sectionc}{1}
\setcounter{subsectionc}{0}\setcounter{subsubsectionc}{0}\noindent
        {\tenbf\thesectionc. #1}\par\vspace{5pt}}
\renewcommand{\subsection}[1] {\vspace{12pt}\addtocounter{subsectionc}{1}
        \setcounter{subsubsectionc}{0}\noindent
        {\bf\thesectionc.\thesubsectionc. {\kern1pt \bfit #1}}\par\vspace{5pt}}
\renewcommand{\subsubsection}[1] {\vspace{12pt}\addtocounter{subsubsectionc}{1}
        \noindent{\tenrm\thesectionc.\thesubsectionc.\thesubsubsectionc.
        {\kern1pt \tenit #1}}\par\vspace{5pt}}

\newcounter{appendixc}
\newcounter{subappendixc}[appendixc]
\newcounter{subsubappendixc}[subappendixc]
\renewcommand{\thesubappendixc}{\Alph{appendixc}.\arabic{subappendixc}}
\renewcommand{\thesubsubappendixc}
        {\Alph{appendixc}.\arabic{subappendixc}.\arabic{subsubappendixc}}
 
\renewcommand{\appendix}[1] {\vspace{12pt}
        \refstepcounter{appendixc}
        \setcounter{figure}{0}
        \setcounter{table}{0}
        \setcounter{lemma}{0}
        \setcounter{theorem}{0}
        \setcounter{corollary}{0}
        \setcounter{definition}{0}
        \setcounter{equation}{0}
        \renewcommand{\thefigure}{\Alph{appendixc}.\arabic{figure}}
        \renewcommand{\thetable}{\Alph{appendixc}.\arabic{table}}
        \renewcommand{\theappendixc}{\Alph{appendixc}}
        \renewcommand{\thelemma}{\Alph{appendixc}.\arabic{lemma}}
        \renewcommand{\thetheorem}{\Alph{appendixc}.\arabic{theorem}}
        \renewcommand{\thedefinition}{\Alph{appendixc}.\arabic{definition}}
        \renewcommand{\thecorollary}{\Alph{appendixc}.\arabic{corollary}}
        \renewcommand{\theequation}{\Alph{appendixc}.\arabic{equation}}
        \noindent{\tenbf Appendix \theappendixc #1}\par\vspace{5pt}}
\newcommand{\subappendix}[1] {\vspace{12pt}
        \refstepcounter{subappendixc}
        \noindent{\bf Appendix \thesubappendixc. {\kern1pt \bfit #1}}
        \par\vspace{5pt}}
\newcommand{\subsubappendix}[1] {\vspace{12pt}
        \refstepcounter{subsubappendixc}
        \noindent{\rm Appendix \thesubsubappendixc. {\kern1pt \tenit #1}}
        \par\vspace{5pt}}
 
\newcommand{\textlineskip}{\baselineskip=13pt}
\newcommand{\smalllineskip}{\baselineskip=10pt}
 
\def\eightcopyright{\copyright}
 
\newcommand{\copyrightheading}[1]
        {\vspace*{-2.5cm}\smalllineskip{\flushleft
        {\eightrm International Journal of Modern Physics C, #1}\\
        {\eightrm $\eightcopyright$\, World Scientific Publishing
         Company}\\
         }}
 
\newcommand{\publisher}[2]{{\begin{center}\eightrm\smalllineskip
        Received #1\\
        Revised #2
        \end{center}
        }}
 
\def\abstracts#1#2#3{{
        \centering{\begin{minipage}{4.5in}\baselineskip=10pt\eightrm
        \parindent=0pt #1\par
        \parindent=15pt #2\par
        \parindent=15pt #3
        \end{minipage} }\par}}
 
\renewenvironment{thebibliography}[1]                   
        {\ninerm
         \baselineskip=11pt                             
         \begin{list}{\arabic{enumi}.}
        {\usecounter{enumi}\setlength{\parsep}{0pt}
         \setlength{\leftmargin 17pt}{\rightmargin 0pt} 
         \setlength{\itemsep}{0pt} \settowidth          
        {\labelwidth}{#1.}\sloppy}}{\end{list}}
 
\newcounter{itemlistc}
\newcounter{romanlistc}
\newcounter{alphlistc}
\newcounter{arabiclistc}

\newcommand{\fcaption}[1]{
        \refstepcounter{figure}
        \setbox\@tempboxa = \hbox{\eightrm Fig.~\thefigure. #1}
        \ifdim \wd\@tempboxa > 5in
           {\begin{center}
        \parbox{5in}{\eightrm \smalllineskip Fig.~\thefigure. #1 }
            \end{center}}
        \else
             {\begin{center}
             {\eightrm Fig.~\thefigure. #1}
              \end{center}}
        \fi}
 
\newcommand{\tcaption}[1]{
        \refstepcounter{table}
        \setbox\@tempboxa = \hbox{\eightrm Table~\thetable. #1}
        \ifdim \wd\@tempboxa > 5in
           {\begin{center}
        \parbox{5in}{\eightrm\smalllineskip Table~\thetable. #1 }
            \end{center}}
        \else
             {\begin{center}
             {\eightrm Table~\thetable. #1}
              \end{center}}
        \fi}
 
\def\@citex[#1]#2{\if@filesw\immediate\write\@auxout    
        {\string\citation{#2}}\fi                       
\def\@citea{}\@cite{\@for\@citeb:=#2\do                 
        {\@citea\def\@citea{,}\@ifundefined             
        {b@\@citeb}{{\bf ?}\@warning
        {Citation `\@citeb' on page \thepage \space undefined}}
        {\csname b@\@citeb\endcsname}}}{#1}}
 
\newif\if@cghi
\def\cite{\@cghitrue\@ifnextchar [{\@tempswatrue 
        \@citex}{\@tempswafalse\@citex[]}}
\def\citelow{\@cghifalse\@ifnextchar [{\@tempswatrue
        \@citex}{\@tempswafalse\@citex[]}}
\def\@cite#1#2{{$\null^{#1}$\if@tempswa\typeout
        {IJCGA warning: optional citation argument
        ignored: `#2'} \fi}}

\def\pmb#1{\setbox0=\hbox{#1}
        \kern-.025em\copy0\kern-\wd0
        \kern.05em\copy0\kern-\wd0
        \kern-.025em\raise.0433em\box0}

\def\fnt#1#2{\footnotetext{\kern-.3em
        {$^{\mbox{\scriptsize #1}}$}{#2}}}
 
\def\fpage#1{\begingroup
\voffset=.3in
\thispagestyle{empty}\begin{table}[b]\centerline{\footnotesize #1}
        \end{table}\endgroup}
 
\def\runninghead#1#2{\pagestyle{myheadings}
\markboth{{\eightit{\quad #1}}\hfill}{\hfill{\eightit{#2\quad}}}}
 
\font\tenrm=cmr10
\font\tenbf=cmbx10
\font\tenit=cmti10
\font\tenit=cmti10
\font\bfit=cmbxti10 at 10pt
 1
 1
 1

\font\ninerm=cmr9

\font\eightrm=cmr8
\font\eightit=cmti8

\def\qed{\hbox{${\vcenter{\vbox{                          
   \hrule height 0.4pt\hbox{\vrule width 0.4pt height 6pt
   \kern5pt\vrule width 0.4pt}\hrule height 0.4pt}}}$}}

\runninghead{}{}

\begin{document}
\normalsize\textlineskip
{\thispagestyle{empty}
\setcounter{page}{1}
 
\renewcommand{\thefootnote}{\fnsymbol{footnote}} 
\def\bsc{{\sc a\kern-6.4pt\sc a\kern-6.4pt\sc a}}
\def\bflatex{\bf L\kern-.30em\raise.3ex\hbox{\bsc}\kern-.14em
T\kern-.1667em\lower.7ex\hbox{E}\kern-.125em X}
 
\copyrightheading{Vol. 0, No. 0 (1997) 000--000}
 
\vspace*{0.88truein}

\def\OR{\hbox{\vrule height 6pt depth -3pt \vrule height 1pt depth 2pt}
\hskip 2pt}
 
\fpage{1}
\centerline{\bf SLIDING BLOCKS REVISITED:}
\centerline{\bf A SIMULATIONAL STUDY}
\vspace{0.37truein}
\centerline{\footnotesize A. R. DE LIMA, C. MOUKARZEL AND T.J.P. PENNA}
\vspace*{0.015truein}
\centerline{\footnotesize\it  Instituto de F\'{\i}sica, Universidade
Federal Fluminense}
\baselineskip=10pt
\centerline{\footnotesize\it Av. Litor\^anea, s/n - Gragoat\'a}
\centerline{\footnotesize\it 24210-340 Niter\'oi, Rio de Janeiro , Brazil }
\vglue 10pt
\centerline{\footnotesize\it  e-mail: arlima@if.uff.br}
\vglue 10pt
\publisher{(received date)}{(revised date)}
 
\vspace*{0.21truein}

\abstracts{\noindent A computational study of sliding blocks on inclined
surfaces is presented. Assuming that the friction coefficient $\mu$ is a
function of position, the probability $P(\lambda)$ for the block to
slide down over a length $\lambda$ is numerically calculated. Our results
are consistent with recent experimental data suggesting a power-law
distribution of events over a wide range of displacements when the chute
angle is close to the critical one, and suggest that the variation of $\mu$
along the surface is responsible for this. }{}{}

\vspace*{1pt}\textlineskip
\section{Introduction}
\vspace*{-0.5pt}
\noindent
\textheight=7.8truein
\setcounter{footnote}{0}
\renewcommand{\thefootnote}{\alph{footnote}}

The dynamics of rigid bodies sliding on inclined planes under the action of
gravity and possibly friction forces is a very old subject, present in most
basic physics courses. However these systems are still being studied in
order to better understand some of the intriguing properties of granular
materials. Several recent papers\cite{drake90}$^-$\cite{dippel96} present
experimental and computational results that demonstrate the complexity of
friction-related phenomena in specific cases. A recent study~\cite{brito} of
cylinders sliding on a rigid aluminum bar by Brito and Gomes shows that
even this simple system may present unusual features. Measuring the number
$N(\lambda)$ of slidings with length larger than $\lambda$ (see figure (1))
the authors find broad regions that can be very well fitted by a power law
$N(\lambda)/N(0) \sim \lambda^{-B}$, where the exponent $B$ does not seem to
depend on the material or the inclination $\theta$ of the chute, when
$\theta$ is close to the critical angle $\theta_{c}=\arctan \mu$.
Their data are in agreement with the Gutenberger-Richter
law\cite{gutenberger56} for the distribution of earthquakes and with
numerical simulations by Chen, Bak and Obukhov\cite{chen91} exhibiting self
organized criticality. However, no theoretical explanation of this behavior was
advanced.
 
In this paper we present a simple model in order to explain the behavior
observed in the above mentioned experiments. We consider the problem of a
block that slides down, under the effect of gravity and friction forces, on
an inclined surface at an angle $\theta$ with the horizontal. We assume that
friction is due to the existence of many uncorrelated contact points between
the surfaces, and therefore becomes a rapidly varying function $\mu(x)$ of
the block position $x$ on the chute. At $t=0$ the block is set in motion
with velocity $v_0$. If $\theta < \theta_{c}$, where
$\theta_{c}=\arctan{\overline{\mu}}$ is the ``critical angle'', the block
will stop with probability one after some finite displacement $\lambda$. We
have in this case an `avalanche' of size $\lambda$. Our aim is to obtain the
probability $P(\lambda)$ that the block stops at position $\lambda$.

\begin{figure}
\label{fig1}
\epsfysize=6cm
\centerline{
\epsfbox{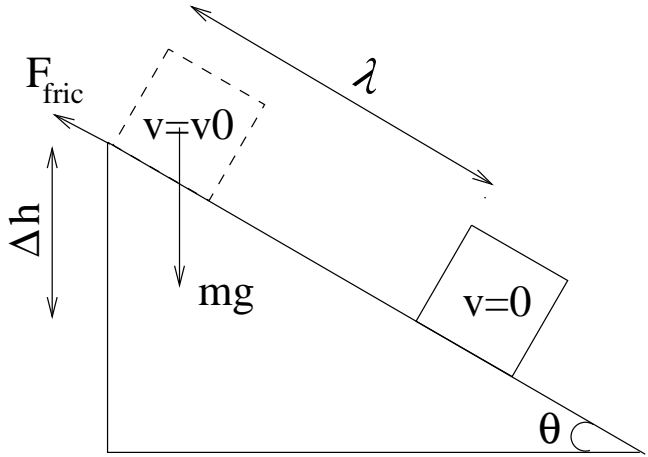}}
\caption{ Schematic diagram of the inclined plane. The blocks is set in motion with velocity $v_{0}$ at $x=0$, and stops at position $x=\lambda$ due to friction.}
\end{figure}

The total energy variation after the slide is
 
\begin{equation} 
\label{eq1} 
\Delta E=mg\Delta h-\frac{1}{2}mv_{0}^{2}\,\,,
\end{equation}

\noindent where $\Delta h=h_f-h_0$ is the vertical displacement, $v_{0}$ the
initial velocity and m the mass of the block. Since energy is dissipated by
friction forces, we also have

\begin{equation} 
\label{eq2} 
\Delta E=\int_{0}^{\lambda}F_{\mu}(x) \,dx = -\int_{0}^{\lambda}mg
\cos(\theta)\mu(x) \,dx 
\end{equation}

\noindent where $\theta$ is the inclination of the plane, $x$ the block
position along the plane, $\lambda (=\Delta x)$ the total displacement and
$\mu(x)$ the local friction coefficient. Equations (\ref{eq1}) and
(\ref{eq2}) admit analytical treatment. The results of our theoretical
approach will be published elsewhere\cite{tbp}.

\section{The Computational Model}

In order to simulate the position-dependent behavior of $\mu$  we introduce
a very simple computational model. This model is based on the following
assumptions: We assume that friction occurs due to randomly scattered
contact points between the two surfaces, and that these contact points are
separated by a characteristic length $a$. The displacement of the block is
discretized in small steps of length $a$, and we represent the ``rugosity'' of both surfaces by means of two binary strings of 0s and 1s. Each bit can be
thought of as representing the average properties of the surface over a
length $a$. If for instance a certain region is more prominent than the
average, the corresponding bit is set to 1, and to 0 in the opposite case.
Thus when the two surfaces are put in contact, only those regions will
contribute to friction which both have a 1 on the corresponding location of
their bit string.

We model the disorder by assigning random strings of bits to both the plane
and the block before each experiment. For computational convenience we set
the block-string length to be 32 bits. Typical plane lengths are on the
other hand $10^5$ bits. The concentration of 1s on the plane- and block-string are
$C_{p}$ and $C_{b}$ respectively.

In Fig.~(2) we show a schematic diagram of a block of length 14 bits with
concentration $C_{b}=0.5$ of ones, sliding over a plane with length $L=39$
bits, with $C_p=0.5$.

\begin{figure}
\label{fig3}
\epsfysize=6cm
\centerline{
\epsfbox{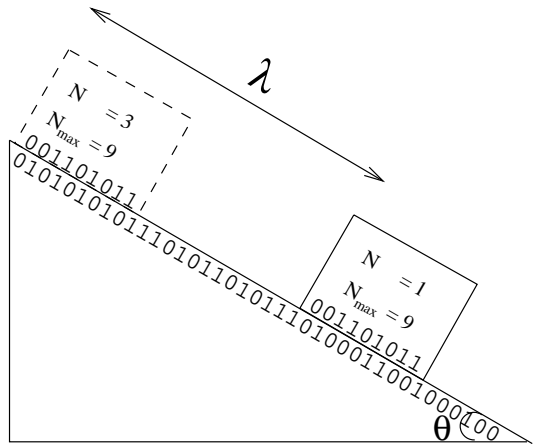}}
\caption{An example of the binary representations of the block
and chute. Only those regions contribute to friction for which both the
block and the plane have a bit set to 1. }
\end{figure}

The local coefficient of friction $\mu(x)$ is defined as

\begin{equation}
\mu(x)=b\frac{N(x)}{N_{max}} 
\end{equation}

\noindent where $N(x)$ is the number of coincident 1s and depends on both
strings, $N_{max}$ is the block length in bits and $b$ is a constant that
can be associated to the strength of each individual contact.

The block motion can be efficiently simulated in the following way:

\begin{enumerate}

\item 

The block starts at $x=0$ with total energy $E_0=\frac12 mv_0^2$.

\item 

We let the block slide over a distance $a$ corresponding to one bit, and its
energy variation is calculated as

\begin{equation} 
\label{eqdeltae}
\Delta E=a F_{\mu}(x) - mga\sin(\theta) = mga \left(
\frac{bN(x)}{N_{max}} \cos(\theta) - \sin(\theta)  \right)
\end{equation}

\item

If the total energy after this change turns out to be zero or negative, the
block has stopped. Otherwise we set $E \to E + \Delta E$ and go to 2.

\end{enumerate}

The critical angle $\theta_c$ can be obtained by setting $<\Delta E>=0$ in
(\ref{eqdeltae}), and satisfies

\begin{equation}
\tan \theta_c =  \overline{\mu} = b C_p C_b
\end{equation}

\begin{figure}[!h]
\label{cgraf1}
\epsfysize=12cm
\centerline{
\rotate[r]{
\epsfbox{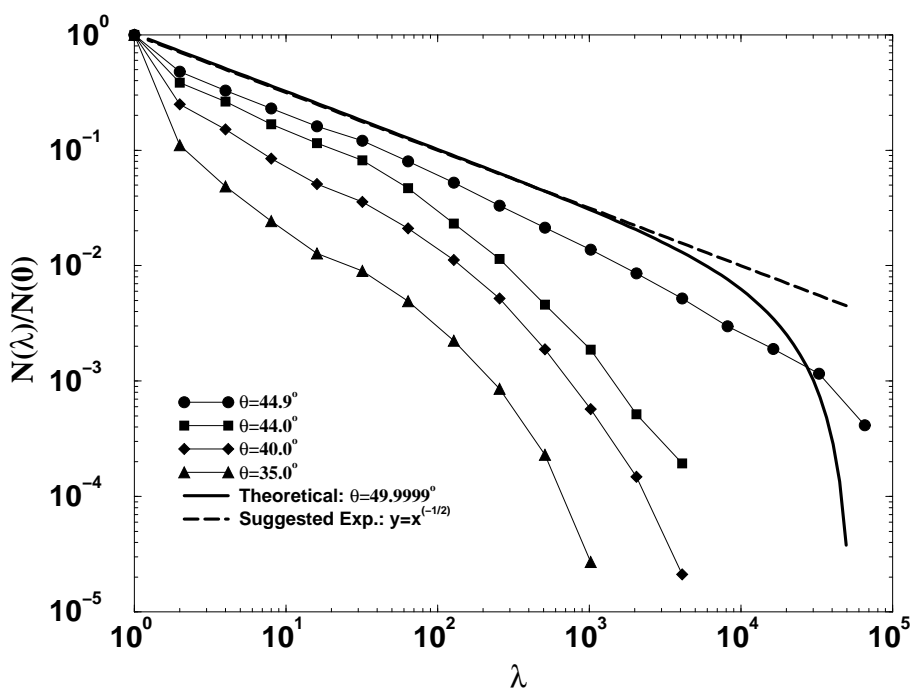}}}
\caption{The accumulated distribution $N(\lambda)$ of slidings
larger than $\lambda$ for several values of the inclination angle $\theta$.
Averages were done over $10^7$ slidings.  The critical angle $\theta_c$ is
$\pi/4$. The straight line $N(\lambda)\sim \lambda^{-1/2}$ shows the
expected behavior for $\theta=\theta_c$.}
\end{figure}

\section{Results} \label{results}

We measure the number $N(\lambda)$ of slidings with size larger than
$\lambda$. We fix $C_{p}=C_{b}=0.5$ and $b=4$ so that $\theta_{c}=\pi/4$. In
Fig. (3) average results are shown for $10^7$ slidings on a plane of maximum
length $10^5$ bits and several values of $\theta \leq \theta_c$.  The
straight line is shown for reference and corresponds to
$N(\lambda)/N(0)=\lambda^{-1/2}$. Our numerical results indicate that, when
$\theta$ is close to $\theta_{c}$, the distribution of avalanches shows a
power-law behavior with exponent $1/2$ over a wide range of sizes. This
exponent has the same value as obtained in experiments\cite{brito}. As
expected on simple grounds, below $\theta_{c}$ the distribution becomes
exponentially decreasing for large sizes and therefore there is a finite
average size. As already mentioned, this problem admits analytical treatment
as well~\cite{tbp}. In Fig.~(3) we show a preliminary theoretical result
corresponding to $(\theta_c-\theta)=10^{-4}$.
 
The average sliding size $\overline{\lambda}$ was also
measured, and the results are presented in Fig.~(4). A power-law $\overline{\lambda}\sim (\theta_{c}-\theta)^{-\nu}$ is obtained with $\nu=1.00\pm0.02$. This value is consistent with $\nu=1$, which is obtained if $\mu$ does not depend on position. In this case $\lambda=\frac{v_{0}^{2}\cos\theta_c}{2g\sin(\theta_c-\theta)}\sim(\theta_c-\theta)^{-1}$, for $\theta$ close to $\theta_c$

\begin{figure}[!h]
\label{cgraf3}
\epsfysize=12cm
\centerline{
\rotate[r]{
\epsfbox{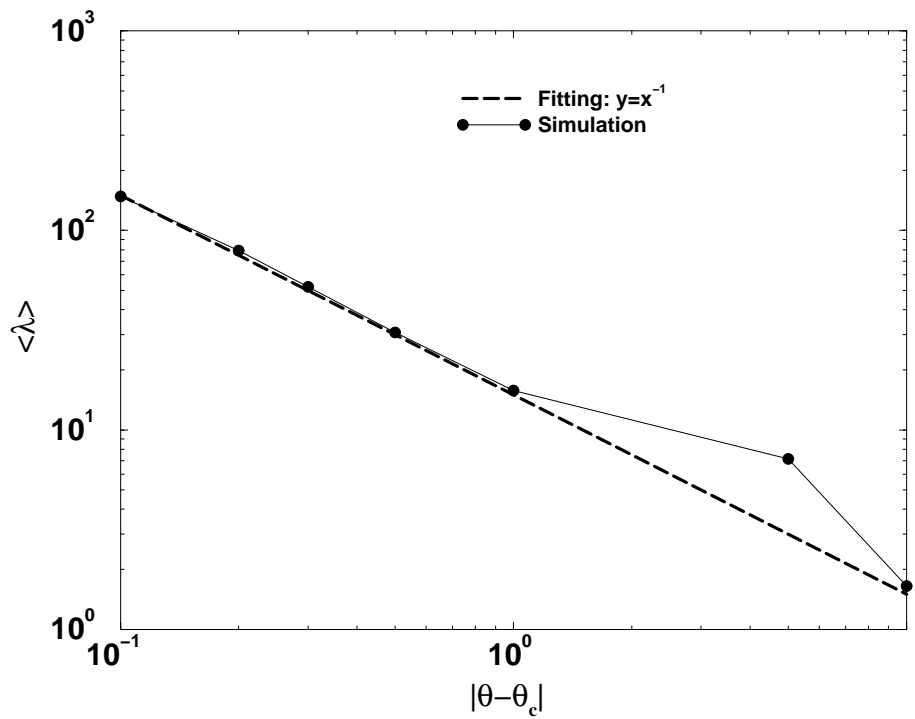}}}
\caption{Mean sliding size as a function of inclination
angle $\theta$.  We see that  $\overline{\lambda}\sim (\theta_c-\theta)^{-\nu}$
holds for  $\theta \to \theta_{c}$.}
\end{figure}

\newpage
\section{Conclusions} 
\label{conclusions}

A very simple model in which the coefficient of friction changes from point
to point on the surface is able to reproduce a power-law behavior in the
distribution of slidings of a block on an inclined chute, as recently
observed in experiments. This holds for values of $\theta$, the inclination
angle, smaller than but very close to $\theta_c$. In this limit, the average sliding size diverges as $(\theta_c-\theta)^{-1}$.

It is possible to do a theoretical study of this problem considering the
random variation of the coefficient of friction. We show a preliminary
result, and a complete study will be published elsewhere~\cite{tbp}.

\bigskip
\noindent{\bf Acknowledgments}
\medskip

This work was par\-tially sup\-ported by Brazil\-ian agen\-cies CNPq, FINEP,
FAPERJ and CAPES. The authors thank H. J. Herrmann for discussions.

}
\end{document}